\documentclass[runningheads]{llncs}
\usepackage{graphicx}
\usepackage{commath}
\usepackage{algorithm, algorithmicx}
\usepackage[noend]{algpseudocode}
\usepackage[caption=false, font=footnotesize]{subfig}

\begin{document}
\title{Evaluating Presidential Support in the Brazilian House of Representatives Through a Network-Based Approach
}
\titlerunning{Evaluating Presidential Support in the House Through Networks}
%
\author{Tiago Colliri\inst{1}\orcidID{0000-0001-9233-4662}}
%
%
\institute{Dept. of Computer Science, University of Sao Paulo, Brazil\\
\email{tcolliri@alumni.usp.br}
}
\maketitle              
\begin{abstract}
Conflicts between the executive and legislative powers are a common, and even expected, characteristic of presidential systems, with some governments being more successful in the activity of obtaining support from the Congress than others. In the case of Brazil, specifically, this factor is considered crucial in the so called ``coalition governments'', with direct positive or negative consequences for the president, in terms of government performance during his (her) term. In this work, we investigate this problem by testing and comparing two different methods for evaluating the government support in the Brazilian House of Representatives. The first method is a more traditional one, being based on roll-call voting data, and measures the presidential support at the \textit{legislators level}. The second method uses a network-based approach, and performs the same type of analysis but at the \textit{parties level}. The obtained results, when applying both methods on legislative data comprising the period from 1998 until 2019, indicate that both methods are valid, with common features being found not only between the results provided by the two of them, but also when comparing their results with those obtained by previous and relevant studies in this field, by using the same type of data but different methodologies.
\keywords{Complex networks \and Centrality measures \and Clustering  \and Legislative voting \and Bills co-sponsorship \and Presidential support \and Political parties.}
\end{abstract}
\section{Introduction}
In presidential systems, the executive and legislative branches of the federal government are independent, with the relationship between the two powers being characterized by periods of more conflict or cooperation. This inter-branch friction has been gaining more attention from researchers in recent years, by using more complex models, in which executive-legislative relations are conceived not necessarily as the interaction between two branches of the government, but as the relationship between the government, political parties and groups of legislators \cite{cheibub201112}. Complex networks, on the other hand, haven been proven suitable tools for capturing and characterizing relationship among data, both physically, topologically and functionally \cite{strogatz2001exploring,colliri2018network,silva12a}. Therefore, not surprisingly, there is an increasingly number of studies using a network-based approach to investigate the relations between the executive and legislative powers. In the Brazilian presidential system, more specifically, building stable coalitions is critical for the elected president to support him (her) on his (her) legislative agenda, in what is currently officially known as ``coalition governments'' among academics, and failing in this activity will bring negative consequences, in terms of government performance. \par
There are currently several works based on networks built from data regarding bills co-sponsorship, with focus on different aspects of the legislative activities, such as: partisanship versus cooperation and collaboration among parties and lawmakers \cite{andris2015,neal2020sign,kirkland2014,moody_mucha_2013,doi:10.1146/annurev.polisci.9.070204.105138,aref2020detecting,waugh2009party}, the formation of interest groups or caucus in the congressmen networks \cite{fischer2019mps,victor2009social}, evaluating government strength or support in the Congress \cite{dal2014voting,prins2006enduring,tsai2020influence,zucco2011distinguishing}, the representation of minority interests in the parliament \cite{epstein2006co}, the relationship between fundraising campaigns and congressmen voting behavior \cite{bursztyn2020brazilian}, and even for corruption prediction purposes \cite{colliri2019analyzing}. In another study \cite{marenco2020time}, which is more directly related to this one, the authors investigated the collaboration and partisanship evolution in the Brazilian House of Representatives, by using minimum spanning tree networks built from roll-call voting data, comprising the period from 2003 until 2008. Among their findings, there is the identification of topological patterns in the lawmakers networks that can be used as indicators of political instabilities over a given period of time. It is also worth mentioning the study \cite{cheibub2009political}, also based on roll-call voting data from Brazilian lawmakers, that investigates whether local and regional interests from the respective states they represent, in the Brazilian federalism system, may influence the legislative decision-making in the House more than the party leaders, at the national level. After analyzing data from 1989 until 2006, to evaluate the impact of local pressures on legislative voting, the authors have concluded that, in fact, political parties play a more central role in the congressmen voting behavior, along with the ability of the president to form stable legislative coalitions to control the House voting agenda. \par
In this paper, we evaluate the presidential support in the Brazilian House of Representatives by using a database built from roll-call voting data from 1998 until 2019, thus covering more than 20 years or legislative activities. Two methods are tested and compared for this end. The first one is based on the ratio of lawmakers votes following the government position, in each session, when compared to the total number of votes, for each period. The second method is a network-based approach, where the political parties and the government are mapped as nodes in the network, and the connections between them are based on their respective vote recommendation similarities, pairwise, in each voting session, over a given period of time.  Therefore, the first method evaluates the presidential support by analyzing the House votes database at the \textit{members level}, while the second method performs the same type of evaluation but at the \textit{parties level}. \par
Regarding the organization of this paper, besides this introduction, we have, in section \ref{sec:method}, the details regarding the database and the methodology used in the study. Then, in section \ref{sec:results}, we present and discuss the obtained results when applying the proposed technique on roll-call voting data from the built database. At the end, in section \ref{sec:remarks}, we close this study by making some final remarks. \par

\section{Materials and Methods}
\label{sec:method}
The methodology used in this study is described below. All network analyses performed in this study are implemented by using the igraph Python package \cite{csardi2006igraph}. \par
\subsection{Database}
The database to be analyzed in this work was built from original data obtained from the Brazilian House of Representatives' official website \cite{website:pcd}. These data comprise the outcome of legislative voting sessions deliberated in the House, in the period from Jan 1, 1998 until July 12, 2019, thus covering more than 20 years of legislative work. The year of 1998 was chosen as the initial one in the built database because this is when information regarding vote recommendations, both from party leaders and from the government, starts to become available in the original data. As preprocessing, a thorough data cleansing was performed in the obtained data, for detecting and fixing possible mistakes, such as duplicated names among lawmakers and misprints. The data for each session comprise the following information: the bill to be considered in the House, the voting date, and the representatives who attended that session. Additionally, the following information is provided for each voter: 
\begin{itemize}
	\item ID (a unique number for each congressperson), 
	\item Full Name, 
	\item Political Party, and 
	\item Vote. 
\end{itemize}
The roll call voting system in the Brazilian House consists of, basically, four types of votes: 
\begin{itemize}
	\item \textit{Yes}: when the lawmaker approves the bill;
	\item \textit{No}: when the lawmaker disapproves the bill;
	\item \textit{Abstention}: when the lawmaker deliberately chooses to not take part in the voting;
	\item \textit{Obstruction}: similar to abstention, except that abstention counts for \textit{quorum} purposes, while obstruction does not count for it.
\end{itemize}
Other types of votes from the original data, mainly consisting of different categories for the ``absent'' situation, were not taken into account in our analysis, being discarded during the preprocessing. \par
At the end, we were able to obtain a total of 1,019,845 votes, from 2,900 different legislative sessions occurred during the considered period. However, not all these sessions presented information about vote recommendations from the part of the government, which is required for this study. Therefore, we filtered these data one last time, to leave only the votes and sessions that presented vote recommendations from the government during the period. The final numbers of the dataset analyzed in this work are shown in Table \ref{tab:data}. As one can observe, the final data cover eight different presidencies, from the first term of Fernando Henrique Cardoso (FHC) until the current government, of Bolsonaro, comprising a total of 814,776 votes in 2,271 distinctive legislative sessions. \par  
\begin{table}
\centering
\caption{Summary of the data used in our analysis. The term ``original'' refers to all voting data available in the period, while ``filtered'' refers only to the data from legislative sessions that present vote recommendations from the government, which are the focus of this work.}
\label{tab:data}
\begin{tabular}{|c|l|r|r|r|r|}
	\cline{1-6}
	{} & &\multicolumn{2}{c|}{\textbf{Original}} &\multicolumn{2}{c|}{\textbf{Filtered}} \\
	\cline{1-6}
	{} \textbf{Period} & \textbf{Presidency} &  \textbf{Sessions} &  \textbf{Votes} &  \textbf{Sessions} & \textbf{Votes} \\
	\hline
01-01-1998 -- 12-31-1998 & FHC I     &        93 &    34,849 &            12 &       4,251 \\
01-01-1999 -- 12-31-2002 &FHC II    &       432 &   163,511 &           329 &     127,880 \\
01-01-2003 -- 12-31-2006 &Lula I    &       451 &   146,538 &           378 &     128,433 \\
01-01-2007 -- 12-31-2010 &Lula II   &       619 &   207,484 &           509 &     173,498 \\
01-01-2011 -- 12-31-2014 &Dilma I   &       428 &   139,007 &           328 &     111,090 \\
01-01-2015 -- 12-02-2015 &Dilma II  &       274 &   112,404 &           247 &     101,401 \\
08-31-2016 -- 12-31-2018 &Temer     &       463 &   161,662 &           358 &     125,155 \\
01-01-2019 -- 07-12-2019 &Bolsonaro &       140 &    54,390 &           110 &      43,068 \\
	\hline
& Total    &      2,900 &  1,019,845 &          2,271 &     814,776 \\
	\hline
\end{tabular}
\end{table}
\subsection{Political Parties Networks Generation}
To perform our analyses, we map each political party in the voting sessions as the node in a network, with one additional node representing the government, and the edges between them are created according to their respective vote recommendations similarity, pairwise, on all sessions of the same period. Given that there are currently 33 political parties in Brazil, then the maximum number of nodes in the network obtained from current voting data, in the hypothetical situation where all parties currently occupy seats in the House, will be 34, i.e., the total number of parties plus one additional node to represent the government. Building a network from bills co-sponsorship data is a well known technique, that already has been extensively applied in other related studies \cite{epstein2006co,neal2014backbone,fischer2019mps}. However, most of these works focus on the analysis of roll-call voting data at the House \textit{members level}, and not at the \textit{parties level}, as in the case of this approach. Besides of this, here we are also including a node to represent the government in the networks, which can be considered a novelty in this type of study. \par
A network can be defined as a graph $\mathcal{G}=(\mathcal{V, E})$, where $\mathcal{V}$ is a set of nodes and $\mathcal{E}$ is a set of tuples representing the edges between each pair of nodes $(i,j) : i,j \in \mathcal{V}$. The edges in $\mathcal{E}$ are usually provided in the form of a square matrix $M$, with size equal to the number of nodes in the network and binary values, in case of unweighted graphs. In our case, we start by mapping the parties vote recommendation similarities, in each voting session $s$, as a square matrix $A$, according to: 
\begin{equation}
\label{eq:A}
A_{i,j}(s)= 
\begin{cases}
1, \text{ if party $i$ and $j$ recommended the same vote in session $s$} ,\\
-1, \text{ otherwise} .
\end{cases}
\end{equation}
The values in $A$ are then accumulated in a matrix $W$, considering all voting sessions occurred in a given time interval $T$, in the form of:
\begin{equation}
\label{eq:W}
W_{i,j}(T)=\sum_s A_{i,j}(s) \mid s_t \in T \quad ,
\end{equation}
where $s_t$ returns the time $t$ when session $s$ has happened. In this way, the values in matrix $W$ will, basically, provide us the overall level of political agreement (or disagreement) between each pair of parties $i$ and $j$, based on their respective vote recommendations in all $n$ sessions occurred within the period $T$. Please note that, consequently, the maximum possible value of $W_{i,j}(T)$ will be $+n$, in case the party leaders from $i$ and $j$ presented the same vote recommendations for all sessions in the period, while, on the other hand, the minimum value will be $-n$, which means that their party leaders gave different vote recommendations in all sessions of that period.\par  
There are different possible ways of generating a network $\mathcal{G}(T)$ from matrix $W(T)$. One of this ways is, for instance, simply mapping it as a weighted network, using the values in $W(T)$ as the edges weights. Or, as another possibility, one can also map the values in $W(T)$ to binary ones, according to a predefined threshold value. In this study, we opt for generating edges between each pair of vertices $i$ and $j$ based on the following rule:
\begin{equation}
\label{eq:edges}
\mathcal{G}_{i,j}(T) = 
\begin{cases}
1, \textrm{ if } W_{i,j}(T) = \displaystyle \max W_i (T) \\
0, \textrm{ otherwise } . \\
\end{cases}
\end{equation}
This method results in an unweighted graph, in which parties are connected only to the ones that are the most politically aligned to them, in terms of their respective leaders vote recommendations in the House. We decide to proceed this way, by considering only the maximum values in matrix $W(T)$ when generating the edges, mainly for the following reasons: (1) as the resulting networks will be less connected, we believe their topological structures will reflect only the most meaningful information, in terms of the parties political similarities, in each period, (2) it also facilitates the interpretability and visibility of the results, and (3) less dense networks are more sensitive to the specific measures that will be extracted from them later, when evaluating the government support in each presidency. \par
With regards to the time interval $T$, we opt for using a yearly frequency for this end, with $T \in [1998, 2019]$, respecting the initial and final dates of each presidency, as listed in Table \ref{tab:data}, which results in 22 different political parties networks generated. \par  
\subsection{Presidential Support Evaluation}
We evaluate the level of support for each presidency in the House of Representatives through two different methods. The first one is based on the actual legislative voting data for the period, i.e., bills co-sponsorship data, and considers the votes registered by lawmakers and the respective vote recommended by the government, in each session, as it has been already made in other similar studies \cite{prins2006enduring,dal2014voting}. The second method is based on specific measures extracted from the built political parties networks, which, as we described above, are based on data regarding vote recommendations, both from party leaders and from the government. We proceed this way with the objective of not only making comparisons between the two methods, but also with the aim of analyzing both results from the perspective of the main political events happened during the last two decades in Brazil. \par   
In the first method considered, we process all roll-call votes $v$, made by each lawmaker $i$, in each session $s$, and the level of presidential support, for a period of time $T$, is provided by:
\begin{equation}
\frac{\sum_s \abs{f(v)}}{\sum_s \abs{v}} \mid f(v)=\{v_i \mid v_i \in v \land v_i = v_g(s)\} \land s_t \in T \quad ,
\end{equation}
where $v_g(s)$ stands for the vote recommended by the government for session $s$, $s_t$ returns the time $t$ when session $s$ happened, and $\abs{v}$ is the length of array $v$. Hence, in another words, this method provides us with the presidential support in the House measured as the ratio of the number of votes from congressmen following the respective government recommendation in each session, when compared to the total number of votes registered during the period.  \par
For the second method, we extract specific measures from the obtained political parties networks, and assess the government support in the House based on these measures, in each period. The network measures considered for this end are described below: 
\begin{itemize}
	\item \textit{Closeness}: a centrality measure, defines how close a vertex is to all other vertices in the graph \cite{okamoto2008ranking}, being calculated as the reciprocal of the sum of the length of the shortest paths between the node and all other nodes in the graph \cite{sabidussi1966centrality}. In the context of this study, a higher closeness centrality for the node representing the government in the political parties network will indicate a higher level of House support for that presidency. 
	\item \textit{PageRank}: also a centrality measure, uses an algorithm originally developed for ranking the importance of website pages in the World Wide Web \cite{page1999pagerank}. In this case, a higher PageRank score for the government, in the political parties networks, indicates more support in the House for that presidency.
	\item \textit{Hub Score}: also known as the HITS algorithm \cite{chakrabarti1998automatic}, is based on the idea that hubs with higher scores represent nodes with links to many other nodes, and a higher authority score is given to nodes that are linked to many different hubs. This indicator will be used in the same manner that the centrality measures described above, with higher scores for the government's node, in the political parties networks, indicating a higher support in the House.
	\item \textit{Network Density}: describes the portion of the potential connections in a network that are actual connections or, in another words, the total number of links over the maximum possible number of links \cite{barabasi2016network}. Differently from the others above, which are analyzed from the node level, this measure is extracted for the whole network, and has already been used in other similar studies, mainly for assessing and comparing connections within political parties \cite{victor2009social,lietz2014politicians,verweij2012twitter}. In this work, we used it as an indicator of how connected are the political parties and also the government in the built networks, for each presidency, with less connections indicating a more divided and fragmented House, hence resulting in more difficulties for the president in obtaining bills approval support.    
\end{itemize}
\par

\section{Results and Discussion}
\label{sec:results}
In this section, we present the results obtained when applying the two considered approaches on the built database, comprising real voting data from around 21 years of legislative work in the Brazilian House of Representatives. \par
\begin{figure}
	\centering
	\subfloat[1998 (FHC I - PSDB)]{%
	\includegraphics[width=.33\linewidth]{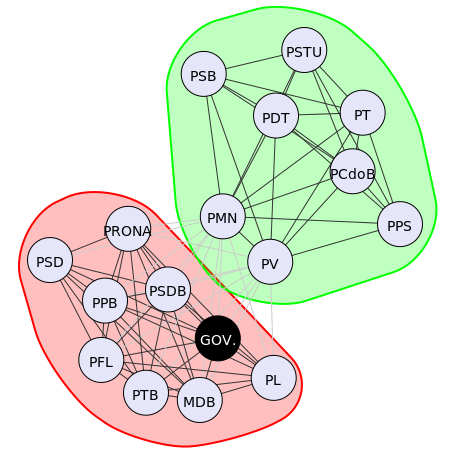}}\hfill
	\subfloat[2002 (FHC II - PSDB)]{%
	\includegraphics[width=.33\linewidth]{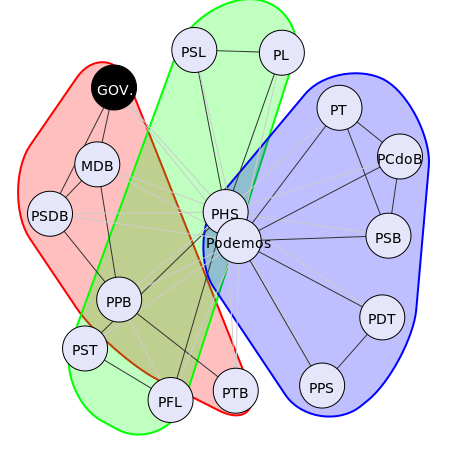}}\hfill
	\subfloat[2004 (Lula I - PT)]{%
	\includegraphics[width=.33\linewidth]{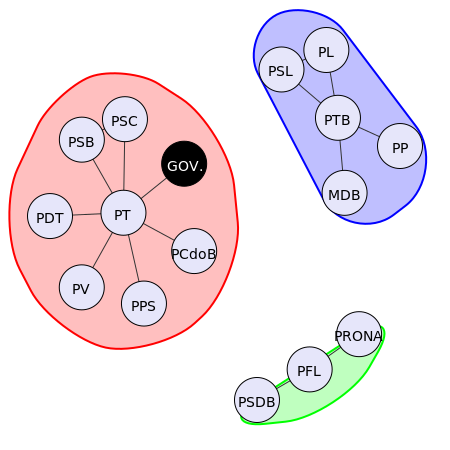}}\hfill
	\subfloat[2009 (Lula II - PT)]{%
	\includegraphics[width=.33\linewidth]{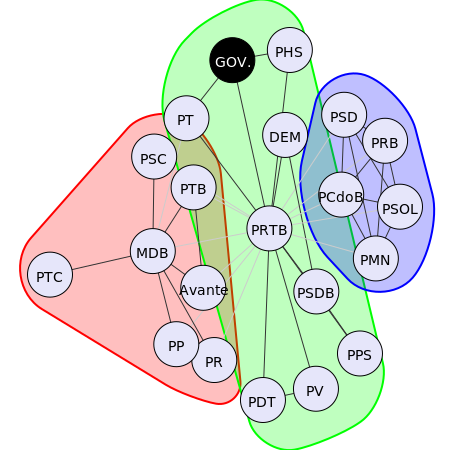}}\hfill
	\subfloat[2012 (Dilma I - PT)]{%
	\includegraphics[width=.33\linewidth]{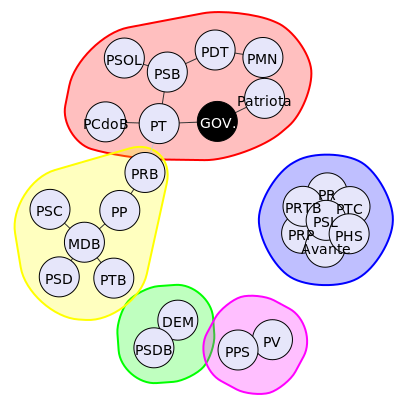}}\hfill
	\subfloat[2014 (Dilma II - PT)]{%
	\includegraphics[width=.33\linewidth]{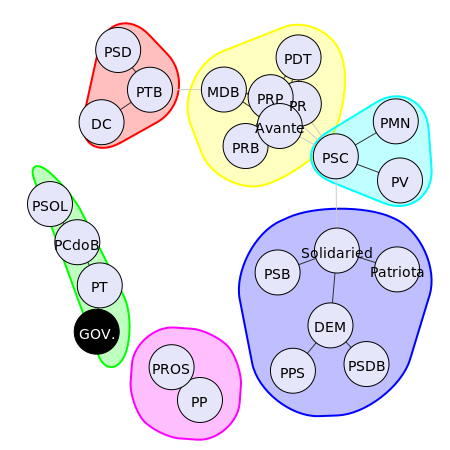}}\hfill
	\subfloat[2016 (Temer - MDB)]{%
	\includegraphics[width=.37\linewidth]{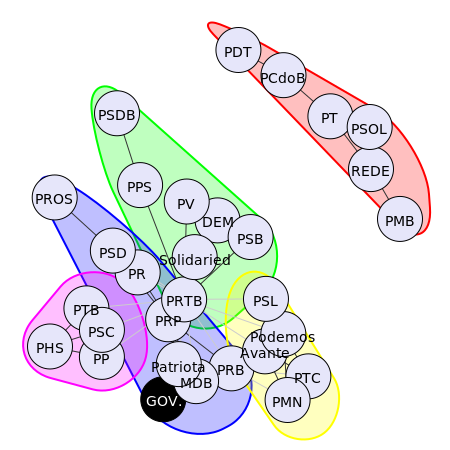}}\hfill
	\subfloat[2019 (Bolsonaro - PSL)]{%
	\includegraphics[width=.4\linewidth]{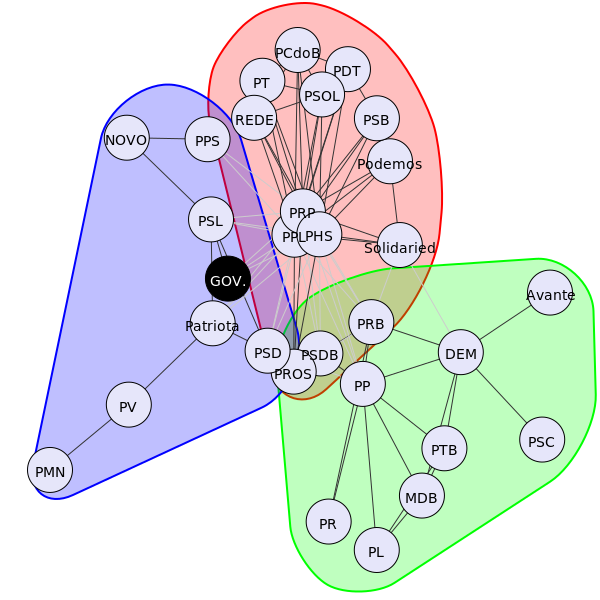}}\hfill
	\caption{Examples of the political parties networks generated in our analysis, based on the party leaders vote recommendations in the House of Representatives, in each year. The communities are detected using the fastgreedy algorithm. The vertex representing the government is highlighted in black.} 
	\label{fig:gov_nets}
\end{figure}

We start by showing, in Fig. \ref{fig:gov_nets}, some examples of the political parties networks generated by the network-based approach, for each year. To help in the analysis, we have also applied the fastgreedy algorithm \cite{brandes2001faster} to detect communities in the networks. The node representing the government is highlighted in black. As one can observe, as a general rule in these networks, the position of the government's node is highly influenced by the party currently in power. In the first graph, for instance, the presidency was from PSDB party, which results in the government's node being directly connected to this party in the network. This same feature can be observed in the graphs from other periods as well, with the node from the government always being closer to the party currently in the presidency, and also to other parties that supported that presidency, in each year. \par
Still regarding Fig. \ref{fig:gov_nets}, we would like to call a special attention to the communities, which, in this case, can be seen as groups of parties that share similar political views, based on the vote recommendations from their leaders in the sessions, for each period. Therefore, for those who are somewhat familiar with the political scenario in Brazil for the last 20 years, it is easy to notice the formation of some sort of ``expected'' aggregations in these networks topological structures. Historical left wing parties, for instance, such as PT, PCdoB. PSB, PDT and PSOL, oftentimes share the same cluster in the networks, in different periods, thus confirming their respective ideological affinities through legislative data. The same observation can also be made for parties historically more associated to right wing political views, such as DEM (former PFL), PL, PP and PSL. These results are overall consistent with those obtained in previous works, using the same type of data but a different methodology \cite{tsai2020influence,zucco2011distinguishing}. However, there are also some unexpected formations resulting from these topological structures, such as the fact that the PT and DEM parties share the same community in the network for 2009, in Fig. \ref{fig:gov_nets}(d), which means that the vote recommendations from their party leaders, in this specific year, were not as different as they were in other years. \par
\begin{figure}
	\centering
	\subfloat[Closeness]{%
		\includegraphics[width=.5\linewidth]{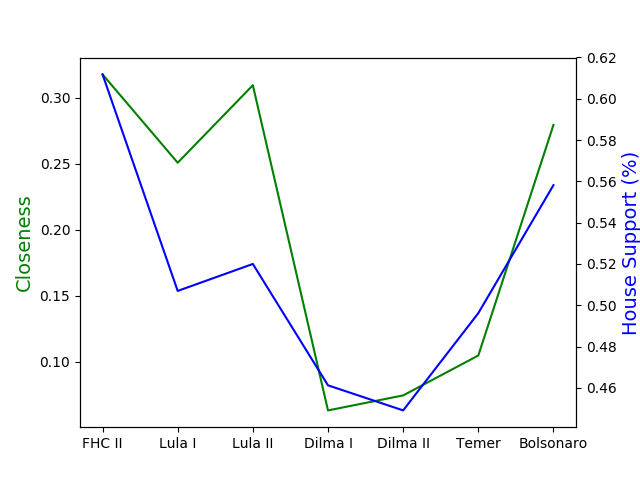}}\hfill
	\subfloat[PageRank]{%
		\includegraphics[width=.5\linewidth]{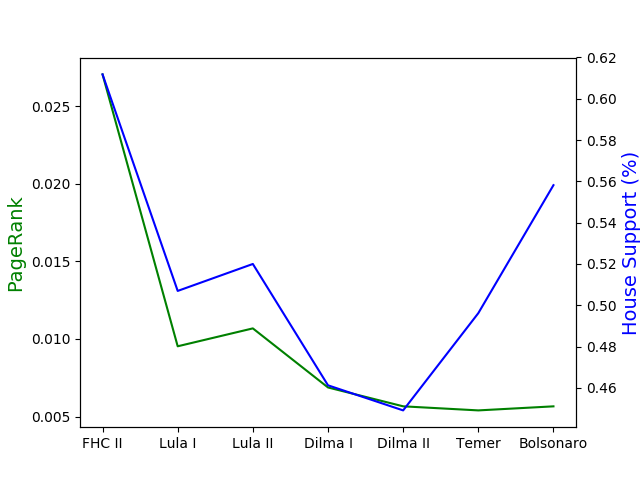}}\hfill
	\subfloat[Hub Score]{%
	\includegraphics[width=.5\linewidth]{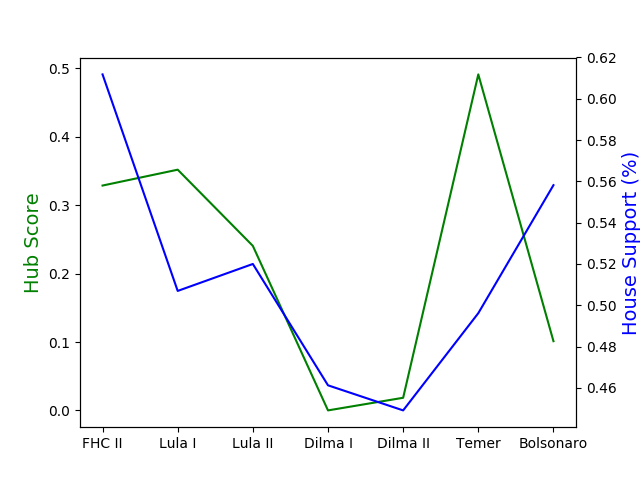}}\hfill
	\subfloat[Network Density]{%
	\includegraphics[width=.5\linewidth]{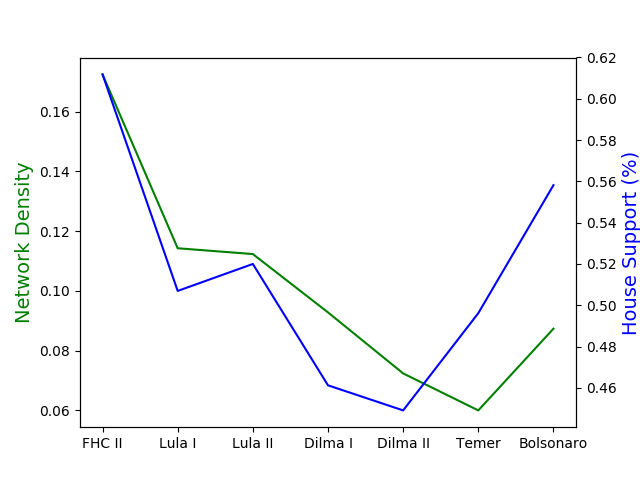}}\hfill
	\caption{Support in the House for each presidency, assessed according to two different methods, for the same period. The first method (in blue) is based on the House members co-sponsorship historical data. The second one (in green) is based on specific measures extracted from the built political parties networks: (a) Closeness, (b) PageRank, (c) Hub Score and (d) Network Density} 
	\label{fig:k_means_ag_cl}
\end{figure}
Let us now proceed to Fig. \ref{fig:k_means_ag_cl}, where we have four plots showing the evolution of the presidential support in the House, evaluated according to the two methods tested in this study. The first method, denoted by the blue color in the plots, is based on co-sponsorship legislative data, considering the ratio of votes registered by congressmen following the respective recommendation from the government, in each session, when compared to the total number of votes recorded within the period of each presidency. The second method tested, represented by the green color in the plots, is based on specific measures extracted from the built political parties networks, with their respective yearly values being averaged, for each presidency. The green line in the first three plots, in Figs. \ref{fig:k_means_ag_cl}(a), \ref{fig:k_means_ag_cl}(b) and \ref{fig:k_means_ag_cl}(c), shows the evolution of each respective measure for the node represented by the government in the networks. Hence, in this form of analysis, the higher the value of each measure in these plots, the higher is the support for that presidency in the House. While, in Fig. \ref{fig:k_means_ag_cl}(d), the density measure is extracted for the whole network and, in this case, less dense networks mean less connections between the nodes and, within this context, indicate a more divided and fragmented House of Representatives, often resulting in more difficulties for the president to build stable coalitions for controlling the House voting agenda. \par
As one can observe, in Fig. \ref{fig:k_means_ag_cl}, although being based on different approaches, both methods overall still present some similar features regarding the evolution of the presidential support in the House during the considered period. Three of the plots show an initial peak for the presidential support evolution, in the FHC II government, followed by a descending behavior, in the Lula presidency, with the only exception for this rule being the analysis based on the Hub Score indicator. This same phenomenon is also observed in previous researches, using the same type of data but different methodologies \cite{cheibub2009political,zucco2011distinguishing}. Another common feature in these analyses is in the fact that all indicators present a descending movement in the first and second terms of the Dilma presidency, indicating that the support in the House was lower during this period, when compared to other ones. This phenomenon makes sense whereas one take into account the fact that the second term of Dilma was interrupted in December 2015, because of the impeachment process, with Temer officially assuming the presidency in August 2016. These results are also in line with those obtained by other researchers, using the same data \cite{marenco2020time}. \par 
Another interesting characteristic that can be noted in Fig. \ref{fig:k_means_ag_cl} is the indication that the second term of Lula had more support in the House than the first one, according to the analyses based on Closeness, PageRank and co-sponsorship voting data. Again, this same feature has also been observed in a previous study \cite{zucco2011distinguishing}, using similar data but a different methodology. There is also the indication that the Bolsonaro presidency has more support in the House than its predecessor Temer, according to the same analysis techniques mentioned earlier plus the one based on Network Density. Additionally, still according to this measure, one can notice that, as Fig. \ref{fig:k_means_ag_cl}(d) indicates, the level of fragmentation among political parties in the House has been increasing since the last term of FHC, reaching its peak during the Temer government, i.e., the bottom in terms of the network density measure, and with a slight pullback in the Bolsonaro presidency. A less dense network, in this context, suggests the incidence of more partisanship and non-cooperation among parties and, consequently, more difficulties for the government to obtain support in the House during the last two decades. This result corroborates as well with those obtained by previous studies, although for US voting data \cite{aref2020detecting,neal2020sign,doi:10.1146/annurev.polisci.9.070204.105138,moody_mucha_2013}. \par 
\begin{figure}
	\centering
	\includegraphics[width=.9\linewidth]{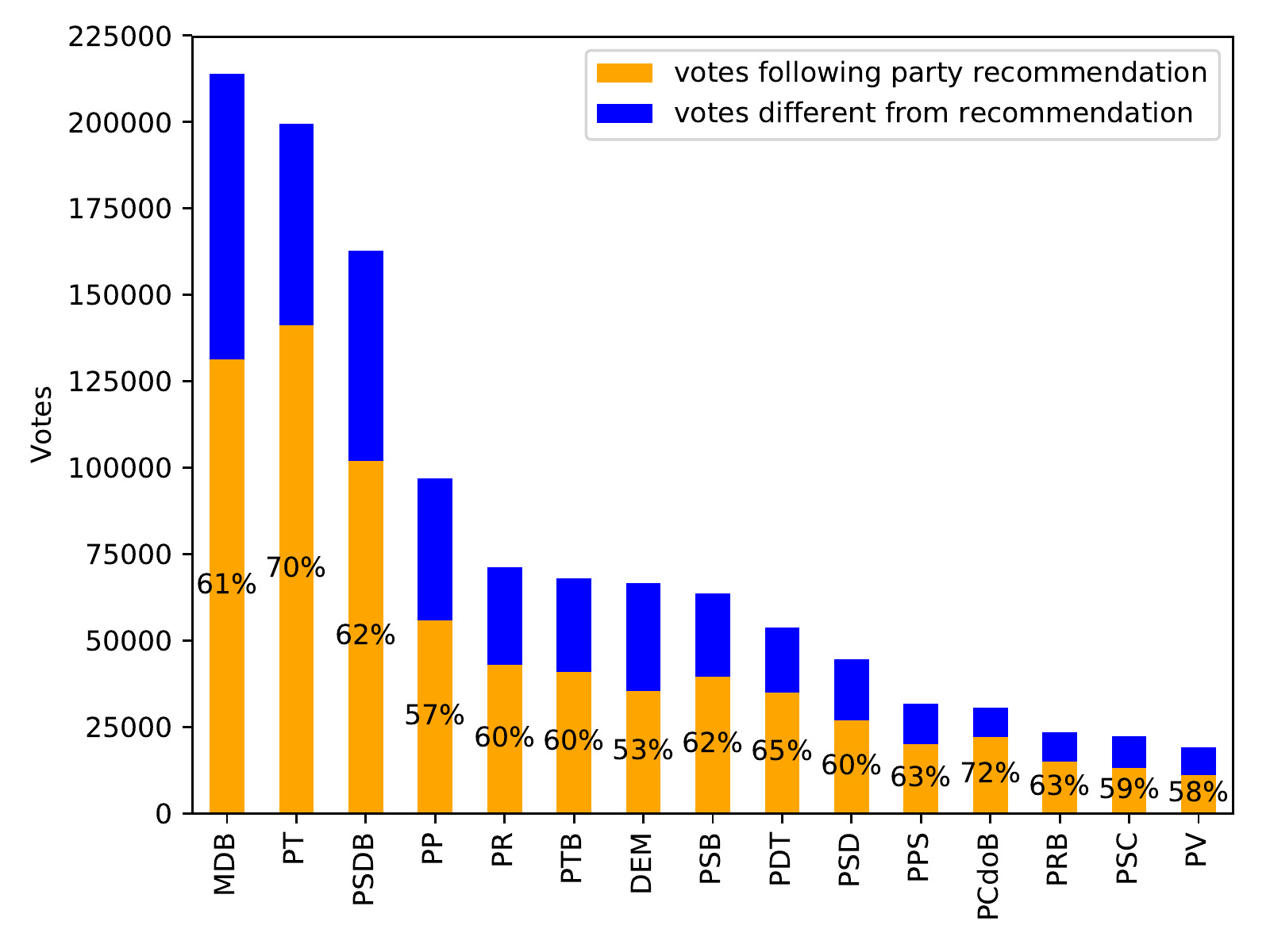}
	\caption{Total votes and loyalty rates for the 15 political parties with highest number of votes in the Brazilian House of Representatives, for the period of 01-01-1998 until 07-12-2019. Only currently active parties are considered for generating this figure.} 
	\label{fig:loyalty}
\end{figure}
We have seen, in Fig. \ref{fig:k_means_ag_cl}, that despite the methods used for evaluating the support for each presidency in the House are different, with one being based on actual voting data and the other being based on the networks built from party leaders vote recommendations, they still present similar features overall, in terms of the evolution of the support in different presidencies. However, the occurrence of such similarities is only possible if lawmakers actually follow the vote recommendations from their respective party leaders, otherwise the analyses resulting from both methods would be completely different. Therefore, to better understand the results, we present, in Fig. \ref{fig:loyalty}, the respective loyalty rates for the main political parties in Brazil, in terms of total number of votes in House sessions during the period considered in this study. Only currently active parties are included in this figure. Overall, the average level of loyalty for the 15 parties is 62\%. The parties with the highest loyalty rates are PCdoB and PT, with 72\% and 70\% of the votes following the recommendation from their leaders during the period, respectively. If we analyze these data also taking into account the total number of votes, the main parties in the House during the period would be MDB, PT and PSDB, with MDB and PSDB having similar loyalty rates, of 61\% and 62\%, respectively. In this aspect, and disconsidering political ideologies, the ideal situation for a government hence would be to receive support, through the formation of coalitions, from political parties that not only have a large number of seats in the House, but that also present a higher rate of loyalty from their members to the parties leaderships, in terms of voting. \par

\section{Final Remarks}
\label{sec:remarks}
In this study, we evaluated the government support in the Brazilian House of Representatives by testing two different methods on data comprising the legislative sessions occurred in the period between 1998 and 2019. The obtained results indicate that the proposed network-based approach is valid, given that, although being different from the more traditional method also tested in this study, which is based solely on co-sponsorship voting data, it was still able to present many common features with the latter, when assessing the support in the House for each presidency during the considered period. Additionally, as this is essentially a graphical approach, there is also the possibility of taking advantage of this as a supplementary resource for both enriching and facilitating the analysis, as we have demonstrated in Fig. \ref{fig:gov_nets}, by presenting the networks generated by the model. Moreover, we also found that most of the results obtained in this work, regarding the ideological similarities between political parties and the evolution of the government support in the House, in each presidency, are in line with the ones achieved by previous relevant studies in this field, based on the same type of data but using different methodologies. \par
As future works, we plan to further explore the built database by making additional analyses regarding the political parties and presidential support. Among these possibilities, we can mention: to verify the existence of a possible correlation between presidential support and the number of government-sponsored bills that were approved in the House, in the same period; to perform analyses based on each type of bill, to find out which types of legislation are more likely to receive support from the members of the House; and also to make a more detailed study regarding the formation of interest groups (or caucus) among the congressmen and/or their respective parties. \par

%
%
%
\bibliographystyle{splncs04}
\bibliography{references}

\end{document}